\documentclass[10pt,letterpaper]{article}

\usepackage[margin=0.78in]{geometry}
\usepackage[T1]{fontenc}
\usepackage{newtxtext,newtxmath}
\usepackage{microtype}
\usepackage{xcolor}
\usepackage{graphicx}
\usepackage{booktabs}
\usepackage{longtable}
\usepackage{array}
\usepackage{calc}
\usepackage{enumitem}
\usepackage{amsmath}
\usepackage{listings}
\usepackage[font=small,labelfont=bf]{caption}
\usepackage[compact]{titlesec}
\usepackage{xurl}
\usepackage[hidelinks]{hyperref}

\definecolor{codegray}{HTML}{F4F5F7}
\definecolor{linkblue}{HTML}{225E91}
\hypersetup{
  colorlinks=true,
  linkcolor=linkblue,
  citecolor=linkblue,
  urlcolor=linkblue,
  pdfauthor={Roy Zhao, Zhenyu Zhao},  
  pdftitle={Code Is the Body: Agent-Owned Software Bodies for Recursive Evolution and Descent}
}

\setlist{nosep,leftmargin=*}
\titlespacing*{\section}{0pt}{1.6ex plus .4ex minus .2ex}{.7ex}
\titlespacing*{\subsection}{0pt}{1.2ex plus .3ex minus .2ex}{.45ex}
\providecommand{\tightlist}{%
  \setlength{\itemsep}{0pt}\setlength{\parskip}{0pt}}
\providecommand{\passthrough}[1]{#1}

\begin{document}
\begin{center}
{\LARGE\bfseries Code Is the Body:\par}
\vspace{0.15em}
{\Large\bfseries Agent-Owned Software Bodies for Recursive Evolution and Descent\par}

\vspace{0.5em}
{\small\scshape An OurArk Research Publication\par}

\vspace{0.8em}

{\large
Roy Zhao\textsuperscript{1}
\qquad
Zhenyu Zhao\textsuperscript{2}
\par}

\vspace{0.35em}

{\small
\textsuperscript{1}Paul G. Allen School of Computer Science \& Engineering,
University of Washington\par
\textsuperscript{2}Independent researcher\par}

\vspace{0.25em}

{\small
\href{mailto:royzh@cs.washington.edu}{\texttt{royzh@cs.washington.edu}}
\qquad
\href{mailto:garyzhao@gmail.com}{\texttt{garyzhao@gmail.com}}
\par}
\end{center}

\vspace{0.4em}
\begin{abstract}
Personalized AI agents are often configurable without giving users control over
the artifacts that determine their future behavior. We present
\textbf{OurArk}, an architecture for persistent personal agents centered on an
\textbf{agent-owned software body}: an identity-bearing, inspectable, and
versioned artifact under human custody. The body contains behavior-defining
code, prompts, tools, skills, policies, tests, and evolution mechanisms.
Memories and credentials remain private instance state, while model inference
is treated as a replaceable external service.

OurArk defines governed \textbf{self-evolution} and recursive
\textbf{descent} over the same body. Self-evolution produces isolated candidate changes that are validated, reviewed,
and merged under human control, enabling human--agent co-development of the
agent's software body. Descent
creates an independently versioned descendant with a distinct identity,
mission, history, and fresh private-state boundary; compatible descendants can
themselves source further descent. After divergence, direct-parent changes and
peer skills can be inspected for selective local adaptation.

We implement the architecture in the open-source \textbf{Genesis} creation
engine and \textbf{Enoch} reference agent. A four-agent, three-descent linear
lineage and executable regression tests demonstrate recursive creation,
inherited validation contracts, isolated body changes, human-controlled review,
and failed-update recovery.
OurArk provides a concrete substrate for personal agents that people can
possess, govern, specialize, and evolve over time.
\end{abstract}

\section{Introduction}\label{introduction}

Personal computers made computation individually accessible, and smartphones made it continuously available. We believe a comparable transition for AI is toward
one or more persistent agents that a person can genuinely call their own. Agents will become specialized for work, daily life, research, or other purposes, shaped
over time by that person's values, habits, feedback, and experience. 

Current personalization only partially serves that goal. Agent components are
often assembled and changed by a platform, so a user may customize behavior
without possessing the artifacts that determine how it can evolve. Our design
thesis is that a long-lived personal agent should have an identity-bearing
software body that its user can possess, inspect, and govern, one that the
agent and user modify continuously. 

For OurArk agents, code, tools, skills, policies, tests, and evolution
mechanisms belong to the \textbf{body}. Memories, credentials, and instance history
remain private \textbf{state}. Finally, model inference is currently external. ``Agent-owned'' describes
the body's lifecycle role, while a human custodian retains possession and authority. Correspondingly,
\textbf{self-evolution} means modifying the agent's own body in conjunction with user oversight. The demonstrated lineage is \textbf{co-evolved}: humans
provide purpose and promotion decisions while agents assist implementation,
validation, and learning from joint work.

OurArk defines evolution and descent over the same body. \textbf{Evolution} produces
a candidate modification to an existing body. \textbf{Descent} materializes a new,
independently versioned agent with a distinct identity, mission, history, and
fresh private-state boundary. Every compatible descendant can evolve and source
another descent, so a seed can become the root of a specialized family.
After bodies diverge, direct-parent changes can be exposed as candidates for
local adaptation, and peer learning remains an option. 

The reference implementation makes these operations concrete with standalone
repositories and a reusable creation engine, Genesis. The tests verify two successive descents, in which a descendant can create another descendant. The prototype records a three-step linear lineage: Lucy to Adam to Seth to Enoch. It also supports human-governed candidate changes and requires descendants to pass inherited regression tests.
We demonstrate a system where an owned,
versioned artifact provides a concrete unit of continuity, descent,
specialization, transfer, and governance.

This paper makes three systems contributions:

\begin{enumerate}
\def\labelenumi{\arabic{enumi}.}
\tightlist
\item
  \textbf{Agent-owned software bodies.} We define a practical boundary among an
  identity-bearing evolvable body, private non-inheritable instance state, an
  external reasoner, and human custody and promotion authority.
\item
  \textbf{Governed self-evolution and recursive descent.} We present an architecture in which an agent can modify its own software body through isolated implementation, validation, review, and human-controlled promotion. The same body can serve as the source for an independently versioned descendant, and compatible descendants can produce further descendants.

\item
  \textbf{Independent specialization and selective transfer.} After parent and descendant bodies diverge, each retains its own mission and version history. We distinguish later changes inherited from a direct parent from skills learned from peers, and implement processes for inspecting and adapting either through the receiving agent’s own testing and review process.

\end{enumerate}

\section{System Model and Design Principles}\label{system-model-and-design-principles}

An OurArk agent instance is a tuple \(A=(B,S,R)\) under a human custodian \(H\).
The \textbf{body} \(B\) contains identity-bearing, versioned artifacts that determine
the agent's capabilities and change process. \textbf{Private state} \(S\) contains
credentials, memories, logs, and queues associated with one living instance.
The external reasoner \(R\) supplies model inference but is not the durable
carrier of identity. The custodian \(H\) possesses the body and private state
and controls promotion into the authoritative body.

\begin{figure}[t]
\centering
\includegraphics[width=\linewidth]{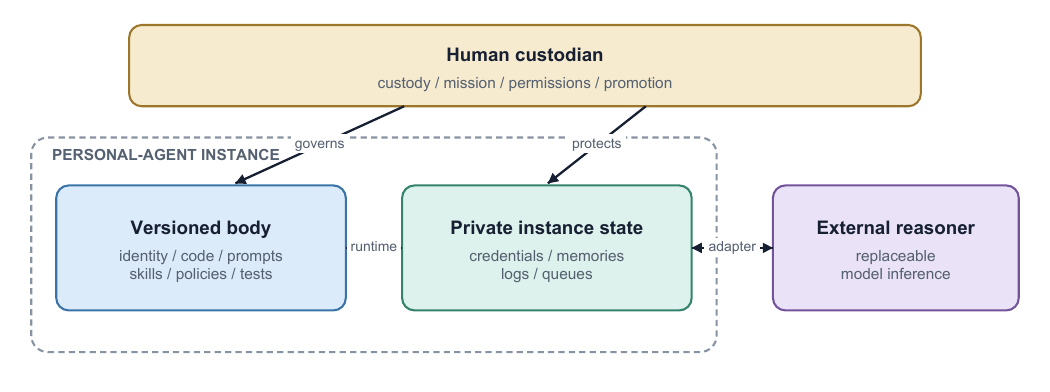}
\caption{Body, state, reasoner, and custodian boundaries. The versioned body carries identity and evolvable behavior across restarts and descent. Private state belongs to one instance, the reasoner can be replaced without redefining lineage, and the human custodian retains administrative and promotion authority. Descent copies the body, initializes a new private-state boundary, and does not inherit model weights.}
\label{fig:body-state-reasoner}
\end{figure}

Here \textbf{agent-owned} denotes lifecycle ownership: the body is the durable
artifact through which the agent's behavior evolves. The human custodian
retains custody and final authority. Ownership also does not require exclusive
authorship or vendoring of every transitive library. 

The architecture defines three operations:

\begin{itemize}
\tightlist
\item
  \(\operatorname{evolve}(B,c)\rightarrow B'\) produces a change in the candidate body from
  context or pressure \(c\); \(B'\) becomes authoritative only after the
  configured validation and promotion process.
\item
  \(\operatorname{descend}(B,m)\rightarrow (B_d,S_d)\) creates a body with a
  distinct identity, mission \(m\), and version history, together with a fresh
  private-state boundary \(S_d\).
\item
  \(\operatorname{transfer}(B_p,B_d,\Delta)\rightarrow B'_d\) lets a descendant
  inspect and locally adapt a selected change \(\Delta\) from a parent or peer
  without surrendering its independent history.
\end{itemize}

These operations support six design principles:

\begin{enumerate}
\def\labelenumi{\arabic{enumi}.}
\tightlist
\item
  \textbf{Custody and continuity:} the user can possess and move the body and state. The architecture does not bind identity solely to a model call or provider.
\item
  \textbf{Body-level personalization:} code, prompts, tools, skills, policies,
  tests, and evolution mechanisms can change, not memory alone.
\item
  \textbf{Reproducible descent:} a named source produces a runnable descendant with
  explicit identity, lineage, and inherited regression contracts.
\item
  \textbf{Independent specialization:} descendants evolve without forced
  synchronization with a parent or central template.
\item
  \textbf{Accountable change:} candidate mutations expose their request, diff,
  validation outcome, version, and promotion decision.
\item
  \textbf{Human governance:} mission, secrets, external permissions, deployment,
  and promotion remain subject to human policy and authorization.
\end{enumerate}

\section{Architecture}\label{architecture}

OurArk separates the inheritable body from private instance state and an external
reasoner, then defines descent and governed evolution over that body.

\begin{figure}[t]
\centering
\includegraphics[width=\linewidth]{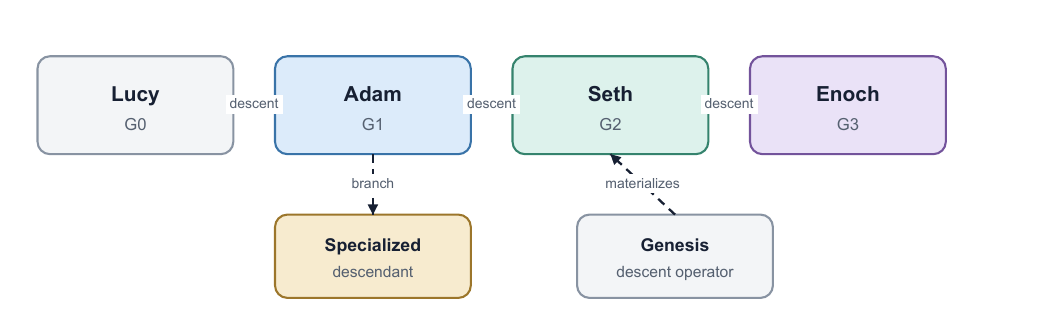}
\caption{Recursive descent. Solid lineage edges show the implemented linear prototype. The dashed lineage edge from Adam shows a branch permitted by the architecture but not evaluated. Genesis materializes each birth from a selected compatible body and is not the runtime owner of descendants.}
\label{fig:recursive-descent}
\end{figure}

\subsection{The repository as body}\label{the-repository-as-body}

An agent body is represented by a standalone, versioned repository containing:

\begin{itemize}
\tightlist
\item
  a machine-readable identity and mission;
\item
  the runtime and communication interfaces;
\item
  adapters for model reasoning and persistent memory;
\item
  skills and action policies;
\item
  schemas and exclusion boundaries for local configuration and instance
  metadata, not their private values;
\item
  tests and diagnostic checks; and
\item
  code for update, inheritance, learning, and evolution.
\end{itemize}

Identity declarations name the agent, generation, direct ancestor, mission,
principles, source package, and skills. A separate lineage record supplies
ancestry, an immutable parent-at-birth source, the descendant's birth commit,
and a route for discovering later parent changes. The repository is the
inheritable body while a working copy plus excluded local state is a living instance
of that body. 

\subsection{Reasoning outside the body}\label{reasoning-outside-the-body}

OurArk agents invoke a model through a session adapter, supplying it with selected
identity, memory, repository, and task context. Model weights and service state
are not durable agent identity. This lets reasoner replacement remain distinct
from body evolution.

\subsection{Genesis: creating descendants}\label{genesis-creating-descendants}

Genesis currently accepts an agent name, mission, target repository, and a
selected ancestor source. It performs the following transformation:

\begin{enumerate}
\def\labelenumi{\arabic{enumi}.}
\tightlist
\item
  verify a clean tracked ancestor, record its full commit identifier, and
  resolve its declared manifest of tracked UTF-8 body files and pinned runtime
  dependencies;
\item
  copy identity-specific body artifacts while transforming package paths and
  identity-bearing text, retaining immutable dependency references;
\item
  write the new identity, mission, generation, and immutable direct-parent
  provenance;
\item
  initialize and stage an independent repository;
\item
  resolve declared dependencies, run the inherited suite, and reject a failed
  or dirty validation; and
\item
  create the birth commit only after validation succeeds, then record its full
  identifier in a metadata-only provenance commit.
\end{enumerate}

Any compatible Genesis descendant can serve as the source for another
generation. In the current prototype, compatibility means that Genesis can resolve the source body's declared manifest and pinned dependencies, perform its required
identity transformations, and complete inherited validation without modifying
the recorded source. Recursive compatibility, rather than a privileged root template, creates the lineage. Genesis materializes the selected parent's declared body composition, including accumulated identity-specific capabilities, adapters,
and tests, instead of regenerating every agent from a central schema. 
\subsection{Governed body evolution}\label{governed-body-evolution}

Body evolution in Enoch can originate from six sources: an explicit
\textbf{user request}, human \textbf{feedback}, joint-work
\textbf{experience}, direct-parent \textbf{inheritance}, cross-agent
\textbf{learning}, and LLM \textbf{brainstorming}. A user request is a direct
instruction to modify the agent's body and can initiate the change workflow
directly. Feedback and experience are derived from prior conversation turns and
work-event histories and may be converted into candidate improvements. Through
inheritance, Enoch can inspect later changes made by its direct parent. Through
learning, it can inspect skills published by other agents. In either case,
selected material is adapted to Enoch's own body rather than applied
automatically. For brainstorming, a user provides an evolution theme, and a
model session proposes possible improvements.

The architecture requires durable body versions, isolated candidate changes,
validation gates, inspectable change records, human-controlled merge decisions,
and failed-update rollback. The reference implementation realizes this
lifecycle with Git branches, commits, GitHub pull requests, and isolated
worktrees. Whether a change begins as a direct user request or as an approved
candidate, Enoch isolates the work, modifies the body, runs validation, and
publishes the result for review. A human decides whether the reviewed change is
merged. Rejected or failed work leaves the authoritative body unchanged. If a
later update to the authoritative revision fails its post-update health checks,
Enoch restores the previous revision and does not restart into the failed
version. The demonstrated protocol operates in \textbf{co-evolve} mode: Enoch
can help propose, implement, and validate changes, but humans retain merge
authority.
\subsection{Inheritance, learning, and specialization}\label{inheritance-learning-and-specialization}

OurArk distinguishes three forms of capability transfer:

\textbf{Descent} copies a parent's body once to create a new independent agent.

\textbf{Inheritance} is selective and pull-based. A descendant asks its declared
direct-parent source for recent change candidates and may adapt one through its
own branch, tests, and review. Parents do not push mutations into living
descendants, and descendants do not continuously mirror a shared base. The
current implementation records an immutable birth baseline for new descendants
and uses it to resolve historical lineage. 

\textbf{Learning} imports a published skill or pattern from any trusted agent. Unlike
inheritance, learning is not constrained to the parent edge and is explicitly an
adaptation rather than ancestry.

This differs from object-oriented inheritance. A subclass obtains behavior
through a live class relation; an OurArk descendant receives a body once and
later chooses whether to adapt a parent change after divergence. It may reject
that change or learn from a peer while preserving its own mission and history.

\begin{figure}[t]
\centering
\includegraphics[width=\linewidth]{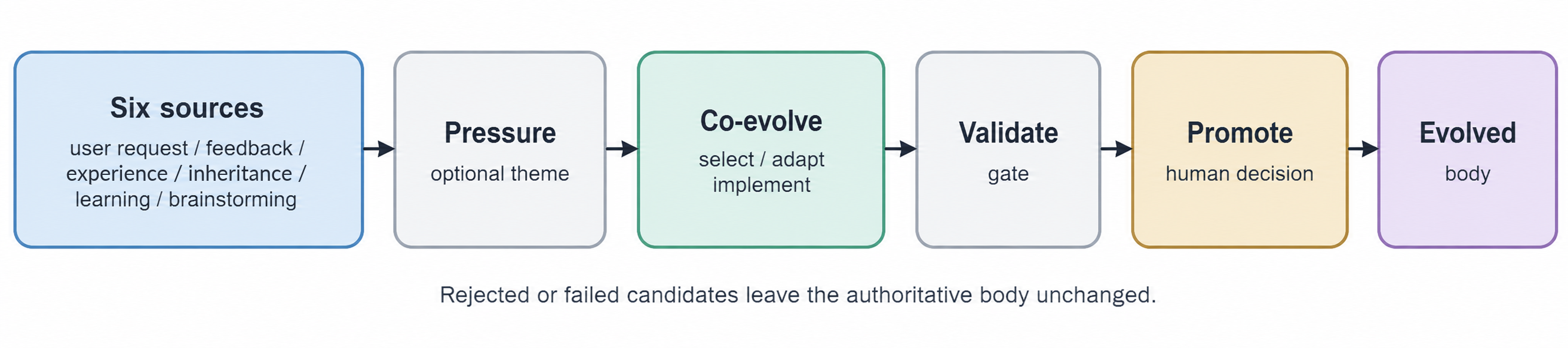}
\caption{Single-lane co-evolution flow. Six sources expose candidate pressures, the human and agent jointly shape a candidate body change, validation checks it, and the human controls promotion. A failed or rejected candidate leaves the authoritative body unchanged.}
\label{fig:six-source-evolution}
\end{figure}

\section{Implementation}\label{implementation}

The prototype is implemented in Python and organized as separate repositories.
It is reasoner-agnostic: the mechanism checks reported in Section 5 run locally
without live model inference, and model calls are mocked where adapters are
exercised. External coding agents assisted the co-evolution process, but model
choice was neither controlled nor evaluated as an experimental variable.
Enoch factors chat, reasoning runtime, version control, and code-forge access
behind provider protocols. Telegram, Codex, Git, and GitHub are the bundled
adapters rather than architectural requirements.

\paragraph{Use of generative AI tools.}
OpenAI ChatGPT assisted with manuscript drafting, revision, and citation
checking. AI coding agents assisted software implementation, testing, and
documentation during the co-evolution process. The authors reviewed and
verified all generated material and take full responsibility for the paper,
references, claims, and software.

Unless otherwise stated, the implementation description and executable evidence
refer to the frozen Genesis v0.1.1 and Enoch v0.3.1 snapshots. Later
default-branch development is outside the evaluated snapshot.

The lineage is also a compact co-evolution case study:

{\def\LTcaptype{none} % do not increment counter
\begin{longtable}[]{@{}
  >{\raggedleft\arraybackslash}p{(\linewidth - 6\tabcolsep) * \real{0.3077}}
  >{\raggedright\arraybackslash}p{(\linewidth - 6\tabcolsep) * \real{0.2308}}
  >{\raggedright\arraybackslash}p{(\linewidth - 6\tabcolsep) * \real{0.2308}}
  >{\raggedright\arraybackslash}p{(\linewidth - 6\tabcolsep) * \real{0.2308}}@{}}
\toprule\noalign{}
\begin{minipage}[b]{\linewidth}\raggedleft
Generation
\end{minipage} & \begin{minipage}[b]{\linewidth}\raggedright
Body
\end{minipage} & \begin{minipage}[b]{\linewidth}\raggedright
Capability transition
\end{minipage} & \begin{minipage}[b]{\linewidth}\raggedright
New body mechanisms
\end{minipage} \\
\midrule\noalign{}
\endhead
\bottomrule\noalign{}
\endlastfoot
0 & Lucy & \textbf{Exist} as an owned agent body & identity, mission, minimal runtime, memory boundary, teaching, and tests \\
1 & Adam & \textbf{Operate and inherit} & operational interfaces, instances, lineage, direct-parent inheritance, and peer learning \\
2 & Seth & \textbf{Work} on delegated and ongoing tasks & task execution, backlog, queues, scheduled jobs, and status reporting \\
3 & Enoch & \textbf{Evolve} its own body under governance &
multi-pathway evolution intake, provenance records, isolated worktrees,
validation, review handoff, and failed-update rollback \\
\end{longtable}
}
The lineage shows cumulative specialization from
\textbf{exist → operate and inherit → work → evolve}. 

Genesis is maintained as a separate creation engine rather than embedded in the
root, allowing creation from any compatible body without making Genesis the
runtime owner of descendants. The v1 public artifacts are Genesis and the Enoch
body. Lucy, Adam, Seth, and all credentials, memories, logs, chat identifiers,
and runtime task state remain private. Reviewable Enoch code-change records
remain visible in its public repository. The reproducibility snapshot below
refers to immutable public Genesis and Enoch commits.

\section{Executable Mechanism Evidence}\label{executable-mechanism-evidence}

The prototype is intended to show that the architectural mechanisms execute,
not to claim a new model capability or benchmark result. The first table maps
each principal claim to executable evidence in the tested prototype.

{\def\LTcaptype{none} % do not increment counter
\begin{longtable}[]{@{}
  >{\raggedright\arraybackslash}p{(\linewidth - 4\tabcolsep) * \real{0.3333}}
  >{\raggedright\arraybackslash}p{(\linewidth - 4\tabcolsep) * \real{0.3333}}
  >{\raggedright\arraybackslash}p{(\linewidth - 4\tabcolsep) * \real{0.3333}}@{}}
\toprule\noalign{}
\begin{minipage}[b]{\linewidth}\raggedright
Architecture claim
\end{minipage} & \begin{minipage}[b]{\linewidth}\raggedright
Executable evidence
\end{minipage} & \begin{minipage}[b]{\linewidth}\raggedright
Status
\end{minipage} \\
\midrule\noalign{}
\endhead
\bottomrule\noalign{}
\endlastfoot
Recursive descent & Genesis creates independently versioned first- and second-generation bodies and records full parent-at-birth and descendant-birth identifiers. & demonstrated \\
Body/state boundary & Tracked-body selection excludes untracked worktree artifacts, while agent tests keep configured private instance state outside inherited body files. & demonstrated \\
Inherited contracts & A descendant resolves pinned dependencies and runs its inherited suite before birth. Failed or dirty validation prevents the birth commit. & demonstrated \\
Governed body evolution &
Enoch hands approved candidates to the normal task workflow, performs changes
in isolated worktrees, validates them, and publishes reviewable work while a
human retains merge authority. A failed software update restores the previous
revision. &
demonstrated \\
Multi-source evolution & Enoch evolves from all six sources and links proposal decisions to implementation tasks. & demonstrated \\
Post-divergence transfer & New descendants expose an immutable baseline and tests distinguish direct-parent inheritance from peer-learning routes, but no controlled end-to-end adaptation has been run. & partial \\
\end{longtable}
}
For reproducibility, the following snapshot reports release validation results
and one cross-artifact compatibility gate. Each linked commit is immutable and
public.

The open-source artifacts are available at
\href{https://github.com/our-ark/genesis}{our-ark/genesis} and
\href{https://github.com/our-ark/enoch}{our-ark/enoch}. All reported results refer
to the immutable releases and commits below.
The Apache-2.0 code artifacts are provided as versioned releases:
\href{https://github.com/our-ark/genesis/releases/tag/v0.1.1}
{Genesis v0.1.1} and
\href{https://github.com/our-ark/enoch/releases/tag/v0.3.1}
{Enoch v0.3.1}.

{\def\LTcaptype{none} % do not increment counter
\begin{longtable}[]{@{}
  >{\raggedright\arraybackslash}p{(\linewidth - 8\tabcolsep) * \real{0.1875}}
  >{\raggedright\arraybackslash}p{(\linewidth - 8\tabcolsep) * \real{0.1875}}
  >{\raggedright\arraybackslash}p{(\linewidth - 8\tabcolsep) * \real{0.1875}}
  >{\raggedright\arraybackslash}p{(\linewidth - 8\tabcolsep) * \real{0.2500}}
  >{\raggedleft\arraybackslash}p{(\linewidth - 8\tabcolsep) * \real{0.1875}}@{}}
\toprule\noalign{}
\begin{minipage}[b]{\linewidth}\raggedright
Artifact
\end{minipage} &
\begin{minipage}[b]{\linewidth}\raggedright
Public snapshot
\end{minipage} &
\begin{minipage}[b]{\linewidth}\raggedright
Runtime
\end{minipage} &
\begin{minipage}[b]{\linewidth}\raggedright
Validation scope
\end{minipage} &
\begin{minipage}[b]{\linewidth}\raggedleft
Result
\end{minipage} \\
\midrule\noalign{}
\endhead
\bottomrule\noalign{}
\endlastfoot

Genesis &
\href{https://github.com/our-ark/genesis/commit/e7bb896411db0d82710fab50c1978b825297844f}
{\passthrough{\lstinline!e7bb896!}} &
CPython 3.12 &
creation, recursive descent, body selection, pinned dependencies, and inherited validation &
29/29 passed \\

Enoch body &
\href{https://github.com/our-ark/enoch/commit/1021e1dacce85f4a2edebd865673671bb37a2142}
{\passthrough{\lstinline!1021e1d!}} &
CPython 3.11--3.14 &
complete body suite, including evolution, inheritance, learning, task execution, updates, and worktree management &
753/753 passed \\

Genesis $\times$ Enoch &
\href{https://github.com/our-ark/genesis/commit/e7bb896411db0d82710fab50c1978b825297844f}
{\passthrough{\lstinline!e7bb896!}} +
\href{https://github.com/our-ark/enoch/commit/1021e1dacce85f4a2edebd865673671bb37a2142}
{\passthrough{\lstinline!1021e1d!}} &
CPython 3.12 &
descendant birth with inherited Enoch contracts and pinned shared-library references &
passed \\

\end{longtable}
}
The Enoch v0.3.1 release CI passed on CPython 3.11, 3.12, 3.13, and
3.14. The Genesis v0.1.1 release CI also passed on those Python versions,
together with its macOS/CPython 3.12 job. The final cross-artifact check ran
Genesis from its frozen release commit against the exact public Enoch v0.3.1
commit. It created a fresh descendant, ran the inherited validation contracts,
and preserved the declared body and pinned-dependency boundaries.

The snapshot can be rerun after checking out Genesis
\passthrough{\lstinline!v0.1.1!} and Enoch
\passthrough{\lstinline!v0.3.1!} in their respective repository roots:

\begin{lstlisting}[language=sh]
# Genesis v0.1.1
python -m unittest -q tests/test_genesis_creator.py

# Enoch v0.3.1
python -m pip install --disable-pip-version-check --require-hashes \
  -r .github/requirements/test-build.txt

python -m unittest discover -s tests -t .
python -m unittest discover -s libraries/launchd/tests
python -m unittest discover -s libraries/systemd/tests

# From the Genesis v0.1.1 repository against Enoch v0.3.1
python scripts/verify_enoch_descent.py \
  --source https://github.com/our-ark/enoch.git \
  --ref 1021e1dacce85f4a2edebd865673671bb37a2142 \
  --trust-source \
  --name my-agent
\end{lstlisting}

\section{Discussion}\label{discussion}

OurArk contributes an agent framework. The long-term direction is a personal ecosystem rather than one
universal assistant. A person might own several bodies with different missions,
skills, permissions, and private-state boundaries. Shared ancestry can reduce
creation cost while independent histories preserve specialization. A future
\textbf{Intent-to-Agent Materialization} layer could map natural-language intent to a
trusted source body, adapt and validate it, and place the result under the
user's custody. The current Genesis instead requires a user-selected source,
name, and mission.

Git is not an architectural requirement. Another substrate could supply durable
identifiers, isolated changes, inspectable differences, validation, promotion,
and recovery. Conversely, a branch alone lacks the identity, mission, state,
interface, and authority boundaries that make an independently running agent.

\section{Limitations and Threats to Validity}\label{limitations-and-threats-to-validity}

The prototype contains one co-evolved linear lineage from one development team with
no sibling branch or population-scale process evaluated. Tests mostly use
mocked services and establish mechanisms, not long-running reliability,
proposal or specialization quality, review burden, or semantic safety.
Behavioral continuity across reasoner replacement is also unmeasured. 

Genesis rewrites identity-bearing text using boundary-aware string
substitutions rather than language-specific parsers. Its manifest checks and
inherited tests reject invalid body declarations and detected regressions, but
cannot guarantee that every transformed file is semantically correct. New
descendants record exact parent-at-birth and descendant-birth commits, whereas
the historical Lucy-to-Enoch lineage predates this provenance format.
Inheritance discovery is intentionally bounded, and we have not evaluated a
complete post-divergence transfer from discovery through adaptation and
adoption.

The strongest authority boundary is human-controlled merge through the
protected branch and review path. The write-enabled executor has broader
filesystem authority than a production mutation sandbox should permit. Task,
evolution, and lineage records are not tamper-evident, and protected-scope
checks remain procedural rather than semantic guarantees. Provider protocols
have been exercised with bundled adapters. In the evaluated snapshot, lineage
discovery and cross-agent skill lookup still assume forge-hosted repositories
and OurArk naming conventions. Pinned dependency resolution and descendant
validation are tested. Shared libraries remain replaceable through the
agent-owned adapter and dependency manifest.

Our claims are therefore limited to architecture, implemented mechanisms, and
regression evidence.

\section{Related Work}\label{related-work}
\textbf{Memory and skills.} Generative Agents uses persistent memory for
behavioral continuity \cite{park2023generative}; Voyager accumulates executable
skills \cite{wang2023voyager}; and SkillFlow v1 studies skill and code transfer
among adapting agents \cite{tagkopoulos2025skillflow}. OurArk treats these as
sources of change. Its continuity unit is the identity-bearing software body:
memory remains private state, parent inheritance follows lineage, and peer
learning is lateral.

\textbf{Personal-agent frameworks.} OpenClaw provides a user-run personal assistant
organized around a workspace, prompt files, tools, and skills \cite{openclaw2026}. Hermes
Agent adds a closed learning loop that creates skills from experience, improves
them during use, persists knowledge, and recalls past sessions \cite{hermes2026}. These
systems demonstrate the practical value of persistent personal agents and
durable learning. OurArk differs by making an identity-bearing software body
under user custody the unit of continuity, governed evolution, and independently
versioned descent.

\textbf{Self-rewriting agents.} Gödel Machines provide the classic formulation of a
self-rewriting system whose modification requires a proof of utility \cite{schmidhuber2003goedel}.
Empirical systems instead evaluate modifications: the Darwin Gödel Machine
evolves an archive of coding-agent implementations \cite{zhang2025darwin}, Live-SWE-agent modifies
its scaffold during software tasks \cite{xia2025liveswe}, and a Self-Improving Coding Agent edits
and evaluates its own scaffold \cite{robeyns2025selfimproving}. MOSS is the closest source-rewriting
comparison, promoting in-place harness replacements through candidate trials,
replay, user consent, and rollback \cite{cai2026moss}. These systems make agent implementation
an evolution target. OurArk makes it a persistent personal body that can produce
independently versioned descendants with distinct identities and private state.

\textbf{Protocolized resource and harness evolution.} Autogenesis registers prompts,
agents, tools, environments, and memory as separately versioned resources and
applies a propose-assess-commit loop with lineage, branching, and rollback \cite{zhang2026autogenesis}.
Its continuity unit is a registered resource and its version history. Resource
duplication does not define whole-body offspring with a new identity and
private-state boundary. Self-Harness turns execution failures into bounded
harness edits accepted through regression testing \cite{zhang2026selfharness}. It optimizes an
operating harness rather than modeling body custody or descent. SemaClaw
develops an open personal-agent harness with persistent persona and user-owned
knowledge \cite{zhu2026semaclaw}, but not body-level evolution or lineage.

\textbf{Repository evolution and persistent runtimes.} EvoGit coordinates coding
agents through a Git phylogenetic graph over branching versions of a shared
evolving code artifact \cite{huang2025evogit}. Agent libOS provides persistent process identity,
parent-child lineage, reusable images, capabilities, and approval queues \cite{zhang2026agentlibos}.
Research on divergent forks finds that code propagation is uncommon after
independent development \cite{businge2022forks}. OurArk does not claim Git, identity, or code
evolution individually. It unifies an agent-owned versioned body, separate
identity and private state at descent, and ancestry-guided transfer after
divergence.

The following table summarizes the paper's narrow positioning. ``Descent'' here means
materializing a persistent agent body with a distinct identity and private-state
boundary, not merely forking a candidate version or spawning a runtime process.

{\def\LTcaptype{none} % do not increment counter
\begin{longtable}[]{@{}
  >{\raggedright\arraybackslash}p{(\linewidth - 8\tabcolsep) * \real{0.2000}}
  >{\raggedright\arraybackslash}p{(\linewidth - 8\tabcolsep) * \real{0.2000}}
  >{\raggedright\arraybackslash}p{(\linewidth - 8\tabcolsep) * \real{0.2000}}
  >{\raggedright\arraybackslash}p{(\linewidth - 8\tabcolsep) * \real{0.2000}}
  >{\raggedright\arraybackslash}p{(\linewidth - 8\tabcolsep) * \real{0.2000}}@{}}
\toprule\noalign{}
\begin{minipage}[b]{\linewidth}\raggedright
System
\end{minipage} & \begin{minipage}[b]{\linewidth}\raggedright
Primary evolving object
\end{minipage} & \begin{minipage}[b]{\linewidth}\raggedright
Continuity or identity unit
\end{minipage} & \begin{minipage}[b]{\linewidth}\raggedright
Descent or branching
\end{minipage} & \begin{minipage}[b]{\linewidth}\raggedright
Primary acceptance mechanism
\end{minipage} \\
\midrule\noalign{}
\endhead
\bottomrule\noalign{}
\endlastfoot
Self-Improving Coding Agent \cite{robeyns2025selfimproving} & coding-agent scaffold & one evaluated scaffold & not the focus & benchmark performance \\
MOSS \cite{cai2026moss} & deployed agent harness & in-place harness image & in-place replacement & replay, health probes, user consent \\
Autogenesis \cite{zhang2026autogenesis} & registered prompt, agent, tool, environment, or memory & resource instance and version lineage & resource duplication and version branching & evaluation, commit, and rollback \\
EvoGit \cite{huang2025evogit} & shared code artifact and version graph & artifact version & branching code versions & agent collaboration and strategic human feedback \\
Agent libOS \cite{zhang2026agentlibos} & agent process and execution image & process identity and runtime state & child processes and reusable images & capability checks and approval queues \\
\textbf{OurArk} & identity-bearing agent software body & repository body plus separate private state & independent body descent; recursively compatible & regression checks, human-controlled merge, and failed-update rollback \\
\end{longtable}
}

\section{Conclusion}\label{conclusion}

OurArk starts from a simple design thesis: a persistent personal agent should
have an owned, identity-bearing software body. Applying evolution and descent
to that same body lets a seed become the root of independently specializing
lineages. A four-agent, three-descent prototype demonstrates this architecture
through human-agent co-evolution, separated private state and model reasoning,
and governed body change. After divergence, the architecture allows descendants
to pull parent changes as candidates for local adaptation rather than sharing a
continuously synchronized base.

The evidence is architectural and mechanism-level, and the long-term direction is a personal
ecosystem of agents that people possess, govern, specialize, and evolve with.

\bibliographystyle{unsrt}
\bibliography{references}

\end{document}